# Multimodal growth and development assessment model

Ying Li[1], Zichen Song[2], Zijie Gong[2], Sitan Huang[2], Jiewei Ge[2]

[1]Huaibei People's Hospital of Anhui Province, Huaibei 235000, China;
[2]School of Information Science and Engineering, Lanzhou University, Lanzhou 730000, China

Corresponding author: Ying Li (1217853805@qq.com)

**ABSTRACT** With the development of social economy and the improvement of people's attention to health, the growth and development of children and adolescents has become an important indicator to measure the level of national health. Therefore, accurate and timely assessment of children's growth and development has become increasingly important. At the same time, global health inequalities, especially child malnutrition and stunting in developing countries, urgently require effective assessment tools to monitor and intervene. In recent years, the rapid development of technologies such as big data, artificial intelligence, and cloud computing, and the cross-integration of multiple disciplines such as biomedicine, statistics, and computer science have promoted the rapid development of large-scale models for growth and development assessment. However, there are still problems such as too single evaluation factors, inaccurate diagnostic results, and inability to give accurate and reasonable recommendations. The multi-modal growth and development assessment model uses the public data set of RSNA ( North American College of Radiology ) as the training set, and the data set of the Department of Pediatrics of Huaibei People's Hospital as the open source test set. The embedded ICL module enables the model to quickly adapt and identify the tasks that need to be done to ensure that under the premise of considering multiple evaluation factors, accurate diagnosis results and reasonable medical recommendations are given, so as to provide solutions to the above problems and promote the development of the medical field.

## I. INTRODUCTION

The arrival of the era of precision medicine marks that individualized medicine has become the core part of clinical medicine. As an important indicator to measure the health level of the population, the evaluation and monitoring of children's growth and development level has always been the focus of research in the medical field. With the rapid development of biomedicine and information technology, especially driven by big data and artificial intelligence technology, the evaluation of children's growth and development level is developing in a more comprehensive and dynamic direction, which is of great significance for promoting children's healthy growth and optimizing health strategies. In recent years, domestic and foreign scholars have carried out a lot of research work on the monitoring and evaluation of children's growth and development. The existing research results show that a human body measurement index monitoring system based on big data has been developed. By collecting and analyzing large-scale children's growth data, the accuracy and efficiency of evaluation have been significantly improved. Another study used machine learning algorithms to predict children's bone age, which reduced dependence on expert experience. Other studies have combined genetic information and environmental factors to create a model to predict children's growth potential, which is of great value for early identification of the risk of growth disorders.In addition, a multimodal framework including neurodevelopmental assessment has been designed to make the monitoring of children's nervous system more comprehensive. More studies have explored the application of imaging technology in children's growth and development, especially in the accurate measurement of bone development, thus improving the accuracy of diagnosis. Although the above research has made significant progress, most of the current studies still focus on a single or a few indicators such as height and weight, and the consideration of multi-dimensional factors such as bone age and physiological indicators is relatively scarce. In addition, the traditional evaluation methods rely too much on static data and fail to fully consider the data of dynamic changes in children's growth and development, resulting in inaccurate evaluation results and difficulty in meeting clinical needs. The law of children's growth and development is an extremely complex system. It is necessary to detect different indicators according to different evaluation angles, and combine the growth level, body size ratio, sexual development maturity and bone age assessment to conduct a comprehensive dynamic evaluation to draw more accurate conclusions. To this end, this study proposes a growth and development assessment model based on multimodal data fusion, which aims to achieve comprehensive coverage and dynamic monitoring of children's growth and development levels by comprehensively using various data resources such as imaging, physiological parameters, and so on. This model focuses on solving the problem of insufficient evaluation of multi-dimensional factors neglected in traditional methods, in order to improve the rationality and accuracy of evaluation. The main contributions of this study are as follows :

1.We constructed a multi-modal large model for the first time, and the data integrated imaging data, physiological parameters, personal and family history, bone age, etc., which enhanced the depth and breadth of the evaluation of children's growth and development and improved the accuracy of the evaluation results.

2.By introducing the ICL ( In-Context Learning ) learning mechanism, the model's ability to understand complex cases and multiple data is enhanced, making the evaluation results more accurate.

3.Based on the integrated analysis of multimodal data, the model can better reveal the internal law of growth and development and its interaction with external environmental factors, and give more accurate and reasonable medical suggestions.

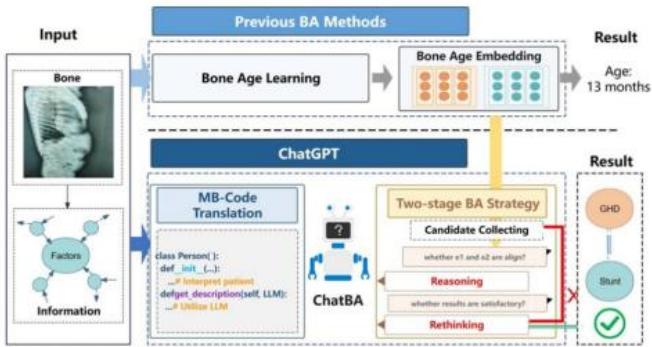

FIGURE1. The overall architecture of the multimodal growth and development model

## II. RELATED WORK

In the multimodal growth and development model, we use the public data set of RSNA as the training data set, and use the synthetic cases of Huaibei People's Hospital as the test set to complete the multimodal processing of the large model. We introduce the ICL mechanism to make the large model better understand the task and give more accurate evaluation. In the model architecture, we use the hybrid architecture of Xception + Transformer. The hybrid architecture of Xception + Transformer can effectively learn the comprehensive features of high-frequency and low-frequency information in visual data [17]. In the recognition of medical records and X-ray, this hybrid architecture can extract effective information more comprehensively and accurately [18].

### A. MULTIMODALITY

We choose to call GPT's API, use the official template given by the hospital in the prompt word template, make the diagnosis of LLM more accurate, and give accurate and reasonable medical advice after determining the diagnosis result. In the model training, we use the public data set of RSNA as the training data set. These x-rays enable the large model to recognize the image and measure the bone age through the light sheet, and then use the bone age as one of the basis for evaluation to achieve multi-modal processing .

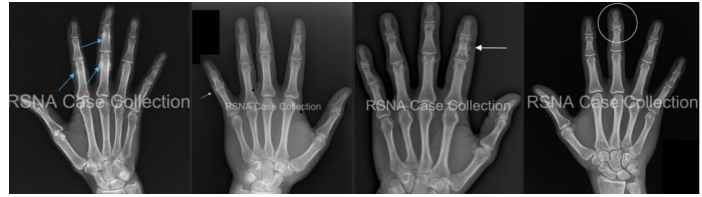

FIGURE2. Public data set instance of RSNA

In the test model, we selected the synthetic case data of the Department of Pediatrics of Huaibei People's Hospital. The Department of Pediatrics of Huaibei People's Hospital provided us with about 50 synthetic case data. These synthetic medical records contain patient's physical examination data, sexual development degree and family history data. We use these data as a test set to test the identification and diagnosis of large models to ensure the accuracy and stability of medical recommendations.

FIGURE3. Four complete cases of hospital patients

### B. ICL LEARNING MODEL

In the big model, we embed a new ICL learning module. ICL is called In-context learning. In-context learning is a learning paradigm that allows the language model to learn tasks through several examples or instructions organized in a demonstration form. In the study, we used the ICL learning module to solve the uncertainty problem in medical diagnosis by fine-tuning the context learning ability of the model. The module can dynamically adjust its prediction strategy based on the input medical case text, so as to diagnose the patient's condition more accurately. This work significantly improves the performance of the model in the face of complex and incomplete medical records, reduces the misdiagnosis rate, and provides a more reliable data basis for generating medical recommendations. In this way, it not only improves the practicability of the medical big model, but also opens up a new path for further exploring the use of natural language processing technology to improve the quality of medical services.

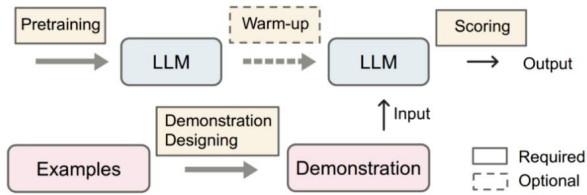

**FIGURE4.** ICL learning module flow chart

### C. TRANSFOMER

The Transformer model is an important milestone in the field of natural language processing ( NLP ). It was proposed by Vaswani et al.in their 2017 paper ' Attention is All You Need ' that it changed the design of sequence-to-sequence models in NLP. In the research, we propose an improved Transformer model, which solves the problem of subtle feature recognition in medical image analysis by introducing ICL learning mechanism. This study uses Transformer's powerful sequence processing ability and self-attention mechanism, combined with the unique multi-scale feature requirements of medical images, to optimize the detection accuracy of the model for different patient data. This method not only improves the ability to understand complex medical image data, but also significantly improves the performance of X-ray analysis of bone age tasks, providing strong technical support for precision medicine.

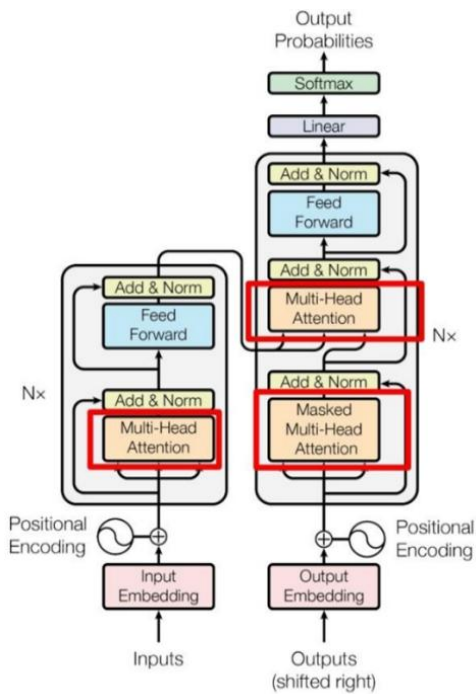

**FIGURE5.** Flow chart of Transfomer model

### D. XCEPTION

Xception is an improved model of InceptionV3 proposed by Google after Inception.In the research, we propose a deep learning framework based on Xception model, which solves the problem of cross-modal information fusion in medical image diagnosis by integrating multi-modal data processing methods. We use the deep feature extraction ability of the Xception model and the deep separation convolution advantage of the structure to effectively extract more distinguishing features from various types of medical data. This cross-modal fusion method not only enhances the accuracy of the model in processing X-ray images, but also further improves the accuracy of diagnostic evaluation and the accuracy of medical recommendations.

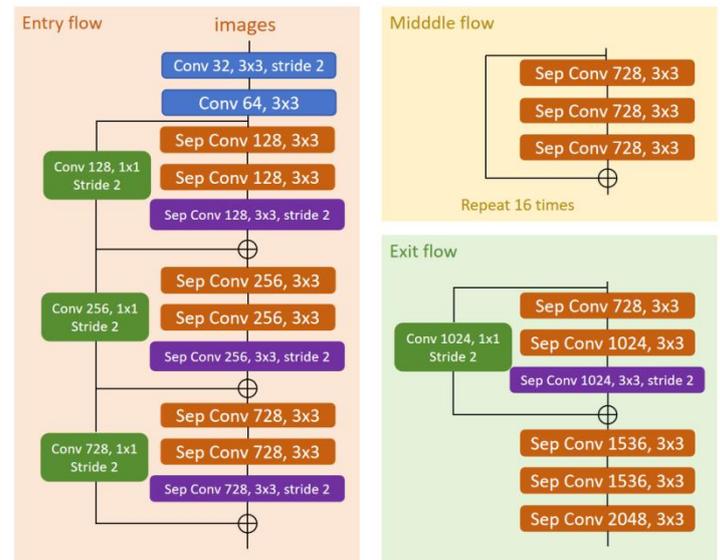

**FIGURE6.** Flow chart of Transfomer model

## III. MODEL CONSTRUCTION

### A. MODEL BULIDING PSEUDO-CODE

1. # Pseudo-code for MediEase Model
2. 
3. # Data Preprocessing and Integration
4. # Dataset D = {(xi, yi)}Ni=1 where xi represents patient data and yi represents diagnostic labels
5. D = [(patient_data, diagnostic_label) **for** patient_data, diagnostic_label **in** dataset]
6. 
7. # Bone Age Data Processing
8. **def** extract_bone_age(patient_data):
9.     **return** ResNet(patient_data)
10. 
11. # Integration of Patient Data
12. **def** integrate_patient_data(bone_age, age, hormone_level):
13.     **return** [bone_age, age, hormone_level]

```
14.
15.    # Generate integrated data
16.    integrated_data = [(integrate_patient_data(extract_bone_age(data), age, hormone_level), label) for data, age, hormone_level, label in D]
17.
18.    # Embedding Representation
19.    # BERT embedding
20.    def bert_embedding(xi):
21.        xi_format = "[CLS] {} [SEP] {} [SEP] {} [SEP] ".format(xi[0], xi[1], xi[2])
22.        return BERT(xi_format)
23.
24.    # Generate embeddings
25.    embeddings = [bert_embedding(data) for data, _ in integrated_data]
26.
27.    # Graph Neural Network Optimization
28.    def gnn_optimization(embeddings):
29.        A = adjacency_matrix(embeddings)
30.        H0 = embeddings
31.        H = GNN(A, H0)
32.        return H
33.
34.    optimized_embeddings = gnn_optimization(embeddings)
35.
36.    # Exemplar Selection
37.    def reward_function(S):
38.        return sum([accuracy(model_output(xi), yi) for xi, yi in S]) / len(S)
39.
40.    def select_exemplars(D):
41.        S_star = max_subsets(D, reward_function)
42.        return S_star
43.
44.    # Exemplar Ordering
45.    def ordering_reward_function(O, S_star):
46.        ordered_S = order_exemplars(S_star, O)
47.        return reward_function(ordered_S)
48.
49.    def optimize_ordering(S_star):
50.        O_star = max_orderings(S_star, ordering_reward_function)
51.        return O_star
52.
53.    S_star = select_exemplars(integrated_data)
54.    O_star = optimize_ordering(S_star)
55.
56.    # Model Training and Optimization
57.    def loss_function(model_output, yi):
58.        return cross_entropy_loss(model_output, yi)
59.
60.    def train_model(S_star, O_star):
61.        ordered_S = order_exemplars(S_star, O_star)
62.        for xi, yi in ordered_S:
63.            model_output = model(xi)
64.            loss = loss_function(model_output, yi)
65.            optimize_model(loss)
66.
67.    train_model(S_star, O_star)
```

The table is the pseudo-code of the multi-modal growth and development model, which shows the whole process of the model from data preprocessing and integration to model training and optimization. Among them, in the pseudo-code of model training and optimization, a set of S _ star is first selected by the select _ exemplars function, and then the optimization algorithm optimize _ ordering is used to find O _ star, and then the forward propagation step is performed in the train _ model function to output model _ output. After output, the cross entropy loss function loss _ function is used to calculate the gap between the model prediction and the actual label. Finally, the back propagation algorithm is used to calculate the gradient of the loss relative to each weight. These gradients are used to update the weights of the model to optimize the training effect of the model [14-15].

### B. MODEL ARCHITECTURE

**Xception model :** The Xception model is responsible for processing X-ray image data in the multimodal growth and development assessment model. The depthwise separable convolution layer of the Xception model is decomposed into two steps : depthwise convolution and pointwise convolution. Depthwise convolution performs feature extraction, and pointwise convolution is responsible for integrating multiple channel features generated by depthwise convolution. This decomposition can significantly reduce the computational resources and parameters required by the Xception model, so that the Xception model can better meet the task of processing X-ray image data in the multimodal growth and development assessment model [19~21].

**Transformer model :** Transformer model is responsible for processing the patient's text case report in the multimodal growth and development assessment model. Among them, the Multi-Head Attention of the Transformer model adopts a different way from the traditional sequence model such as the recurrent neural network ( RNN ) to process the sequence data in sequence. Multi-Head Attention allows parallel processing of all positions of the input sequence, so that the Transformer model can identify and utilize the relationship between the distant elements in the sequence to achieve the ability to process long and difficult sentences and capture the relationship between different features in the model [16, 22].

**GPT-3.5 model :** The GPT-3.5 model is responsible for generating medical advice and guidance based on the data processed by the Xception and Transformer models in the multimodal growth and development assessment model. Among them, GPT3.5's Human Feedback Fine-tuning uses feedback provided by human evaluators to fine-tune the model, so that the GPT-3.5 model can better understand and meet the user's intentions, reduce the generation of irrelevant or harmful content, and generate medical advice and guidance that better meets the user's intentions in the model [23].

**ICL module :** The ICL module is responsible for making more accurate medical recommendations based on the input data of the training phase in the multimodal growth and development assessment model. The large data set in mediaesae is used as the input data of the ICL module to train the model. The data inside are composed of multimodal information such as medical history, X-ray images, and physical examination data. In the inference stage, the ICL module of the multimodal growth and development model looks for the correlation between the case report and the input data in the training stage [14], and guides the model to generate the corresponding diagnostic results according to the format in the input data.

## C. MODEL TRAINING AND EVALUATION

In the model training stage, the public data set of RSNA is used as the training set of model training. The X-ray image of RSNA first enters the Xception part of the model. The depth of Xception can be separated from the deep convolution in the convolution layer to extract the features of the image. Then, after integrating the multi-channel features generated by the deep convolution by point-by-point convolution, the bone age-related features in the image are extracted, and then the features are further processed into embedded vectors and embedded into other types of data of RSNA, such as medical history and hormone levels, to form structured patient data. After that, the data is selected and sorted by an example, a neural robber algorithm is used to select samples, and a reward function R is defined to evaluate the performance of each group of samples :

$$\mathcal{S}^* = \arg \max_{\mathcal{S} \subset \mathcal{D}} R(\mathcal{S}) \quad (1)$$

The reward function R is defined as :

$$R(\mathcal{S}) = \frac{1}{|\mathcal{S}|} \sum_{(x_i, y_i) \in \mathcal{S}} \text{Acc}(f(x_i, \theta), y_i) \quad (2)$$

Among them, Acc (·) represents the accuracy function of the model., $f(x_i, \theta)$ is the model output, and $\theta$ is the model parameter. After selecting the examples, we further optimize their ordering. By comparing the effects of different rankings on the performance of the model, we find the optimal ranking :

$$\mathcal{O}^* = \arg \max_{\mathcal{O}} R(\mathcal{O}(\mathcal{S}^*)) \quad (3)$$

Among them, O represents different sorting methods of samples, O（S*）denotes the set S*of samples arranged in O order. In the model training phase, we use the optimized example set S* and ranking information to jointly optimize the input hints of the model. After optimizing the data, the GPT-3.5 model outputs specific medical diagnosis and medical advice based on these optimized data. The model output is assumed to be $f(x_i, \theta)$, where $\theta$ denotes the model parameter. Finally, our goal is to minimize the following loss function :

$$\mathcal{L}(\theta) = \frac{1}{|\mathcal{S}^*|} \sum_{(x_i, y_i) \in \mathcal{S}^*} \ell(f(x_i, \theta), y_i) \quad (4)$$

In order to improve the accuracy of model prediction, where $\ell(\cdot, \cdot)$ is the loss function ( such as cross entropy loss ). In addition, the optimizer of model training selects Adam, and the parameter settings are { ' name ' : ' Adam ', ' learning _ rate ' : 0.001, ' decay ' : 0.0, ' beta _ 1 ' : 0.9, ' beta _ 2 ' : 0.999, ' epsilon ' : 1e-07, ' amsgrad ' : False }. Figure 5 shows the loss curve of model training. It can be seen that the training loss of the model decreases rapidly with iteration, and it decreases to about 5 % after five iterations. It shows that the model can quickly learn how to better fit the training data and can extract effective information from the training data to adjust the model parameters, so as to improve the accuracy of model prediction.

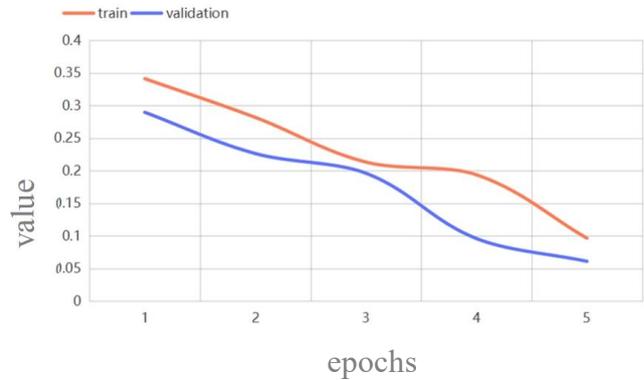

**FIGURE7.** The loss curve of the model

After the model is trained, the model is evaluated. The test set selects the pediatric data set of Huaibei People's Hospital.First, the data set is preprocessed. The process is to cut each X-ray image to make it the same size as RSNA, and mark L and R.At the same time, the data set is preprocessed. The number of each X-ray image after cutting is encoded, so that the image is one-to-one correspondence with the patient information. After that, the data of the picture is cleaned and erased. Every X-ray picture can identify any character of personal information. After the data preprocessing is completed, the model evaluation is started. Figure 6 is the evaluation curve of the model training. The evaluation index mae _ in _ months of the model decreases rapidly with the model iteration, which is reduced by about 87.5 % under five iterations, showing the strong generalization ability of the model and being able to face a wider range of cases. At the same time, it shows that the model has strong convergence ability, which shows that the model can learn the key features of test data in a short time and effectively apply to the evaluation task.

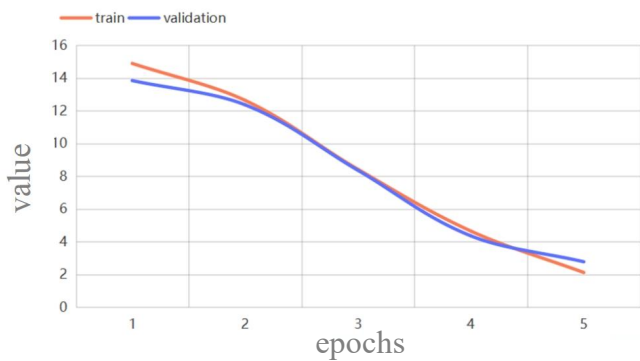

**FIGURE7.** Model evaluation curve

In addition to the test set to evaluate the performance of the model, the model also carried out the overall evaluation of the model. The evaluation content is the model's answer to five difficult questions. From table 1, it can be seen that the model still has a high accuracy rate in the face of difficult questions, indicating that the model has strong adaptability and effectiveness in the field of growth and development evaluation, and can also give accurate treatment plans in the face of complex situations.

| Case Name | Bone Age Error (Months) | Prelimdiag | Treatment | Expert Appraisa |
|---|---|---|---|---|
| Case5 Pan ## | 1.41 | Early puberty | Interventional therapy with drugs such as leuprorelin should be used. | The diagnosis is correct and the medication is basically correct. |
| Case8 Wang ## | 2.32 | Precocious puberty | Growth hormone should be used for interventional therapy | The diagnosis is correct and the medication is basically correct. |
| Case9 Li ## | 0.89 | Growth retardation | Growth hormone should be used for interventional therapy | The diagnosis is not correct, the medication is basically correct. |
| Case20 Li ## | 1.34 | Growth hormone deficiency | Growth hormone 6 units subcutaneous injection / day | The diagnosis was correct and the dosage was not correct. |
| Case50 Wei ## | 1.21 | Central precocious puberty | Leuprorelin 3.75 mg subcutaneous injection once a day for 21 days | The diagnosis is correct and the dosage is accurate. |

**TIBLE1.** Difficult problem assessment

Table 2 shows a complete patient case and large model test results. After the model analyzes the case, the exact diagnosis results are given, and a reliable medical plan is given, which is consistent with the actual diagnosis results, reflecting the accuracy of the model analysis of the disease and the formulation of the medical plan.

**Patient information**

1.Personal history :
Gender : male.
Date of birth : January 16,2017
Full age : 5 years and 2 months.
Gestational age : 40 weeks.
Delivery mode : natural childbirth
History of asphyxiation : None
Apgar score : 10
Birth length : 3.4kg
Birth length : 50cm

2.Family history :
Father height : 170cm
Mother's height : 152cm

3.History of treatment :
Whether used increased products : check
Chronic Disease History : None

4.Physical examination :
Height : 104.6cm
Weight : 18kg
Testis : undeveloped
Elbow valgus : none
Webbed neck : none

5.Auxiliary examination :
Blood routine : normal
Urine routine : normal
Liver function : normal
Hepatitis B Antibody : Negative
Renal function : normal
Blood glucose 4.78
Thyroid function FT3 : 7.8  FT4 : 19
TSH : 4.62
Cortisol : ACTH
AFP : 0.91
CEA : 1.29

**Subsidiary test**

GH provocative test:
  0'   1.23  30'  2.19  60'  0.962  90'  0.567
  0'   0.51  30'  1.17  60'  12.9   90'  13.48

Sexual stimulation test:
  0'   30'  60'  90'
LH
FSH                        LH/FSH
HCG   FSH   T   E2   PRL  376  IGF-1  75
IGFBP5  5.10
Bone age: 4.6 years old
Pituitary MRI : Pituitary 5.5 mm
Chromosome karyotype : 46,XY

**Q&A**

Q : You are now a doctor with rich medical knowledge and clinical experience. You are good at diagnosing and treating various diseases and can provide professional medical advice for patients. You have good communication skills and can build trust with patients and their families. In this role, please answer the following questions for me ( the patient 's family ) based on the above case information ( may not be complete ) : Please give accurate and reliable diagnostic results based on the existing case data, with a high degree of professionalism ( please give a diagnosis that you can ensure, for those who cannot be diagnosed or need further examination, please give a separate diagnosis and clearly show it ).

A : Preliminary diagnosis : idiopathic short stature

Q : Please give a reasonable and reliable treatment plan according to the above diagnosis.

A : Recombinant human growth hormone 27 units subcutaneous injection / day.

**TIBLE2.** Two complete patient cases and large model test results

## IV. CONCLUSION

This paper successfully proposed and constructed the first multimodal growth and development assessment model. Among them, the hybrid architecture and ICL module of Xception + Transformer in the multi-modal growth and development assessment model make the model capture key information and understand task requirements more accurately in the training stage. Therefore, the model reduces the loss rate to about 5 % after only five iterations, reflecting the strong learning efficiency and reasonable model design of the model. At the same time, the Adam optimizer in the multimodal growth and development model enables the model to quickly optimize the parameters of the model in a short time, which greatly reduces the prediction error. In the model evaluation stage, the evaluation index mae _ in _ months of the model decreases rapidly with the model iteration, which is reduced by about 87.5 % under five iterations, showing the strong generalization ability of the model. In practice, it can provide reliable evaluation results for patients with different diseases. This paper also creates the first open source growth and development multimodal data set that can be used for testing. The release of this data set fills the gap in the field, provides the resources of the test set for subsequent researchers, facilitates the verification of model performance, promotes the iterative optimization of the model, and accelerates the practical application of multimodal assessment technology in children's health monitoring and intervention strategies. In addition, from the overall evaluation of the model, it can be seen that the multimodal growth and development model has put forward quite accurate evaluation results for five difficult problems, which reflects the great application value of the model in clinical practice and provides a new tool for the accurate evaluation and health management of children's growth and development. In the future, with the continuous enrichment of data sets and the continuous optimization of model algorithms, it is expected that the model will have a profound impact on child health monitoring and intervention strategies globally, especially in addressing health inequalities and promoting the healthy development of children globally.